\documentclass[10pt,journal,epsfig]{IEEEtran}

\usepackage[dvips]{graphicx}
\usepackage{graphicx}

\usepackage{amssymb}
\usepackage{cite}
\usepackage{amsmath}
\usepackage{algorithm}
\usepackage{algorithmic}
\usepackage{multirow}
\usepackage[table]{xcolor}
\usepackage{subfigure}
\usepackage{makecell}
\usepackage{diagbox}
\usepackage{array}
\usepackage{threeparttable}
\usepackage{graphicx}
\usepackage{caption,epstopdf}
\usepackage{color}

\newcommand{\bm}{\boldsymbol}

\begin{document}

\title{A Derivative-Free Position Optimization Approach for Movable Antenna Multi-User Communication Systems}

\author{Xianlong Zeng, Jun Fang, Peilan Wang, Weidong Mei, and Ying-Chang Liang,
~\IEEEmembership{Fellow,~IEEE}
\thanks{Xianlong Zeng, Jun Fang, Peilan Wang, and Weidong Mei are with the National Key Laboratory
of Wireless Communications, University of Electronic Science and
Technology of China, Chengdu 611731, China, Email:
JunFang@uestc.edu.cn}
\thanks{Ying-Chang Liang is with Center for Intelligent Networking and Communications, University of Electronic Science and
Technology of China, Chengdu 611731, China, Email: liangyc@ieee.org}
} \maketitle


\begin{abstract}
Movable antennas (MAs) have emerged as a disruptive technology in wireless communications for enhancing spatial degrees of freedom through continuous antenna repositioning within predefined regions, thereby creating favorable channel propagation conditions. In this paper, we study the problem of position optimization for MA-enabled multi-user MISO systems, where a base station (BS), equipped with multiple MAs, communicates with multiple users each equipped with a single fixed-position antenna (FPA). To circumvent the difficulty of acquiring the channel state information (CSI) from the transmitter to the receiver over the entire movable region, we propose a derivative-free approach for MA position optimization. 
The basic idea is to treat position optimization as a closed-box optimization
problem and calculate the gradient of the unknown objective function using zeroth-order (ZO) gradient approximation techniques. Specifically, the proposed method does not need to explicitly estimate the global CSI. Instead, it adaptively refines its next movement based on previous measurements such that it eventually converges to an optimum or stationary solution. Simulation results show that the proposed derivative-free approach is able to achieve higher sample and computational efficiencies than the CSI estimation-based position optimization approach, particularly for challenging scenarios where the number of multi-path components (MPCs) is large or the number of pilot signals is limited.   
\end{abstract}



\section{Introduction}\label{section1}
Flexible-position antenna technologies, including movable antenna \cite{zhu2025tutorial} (MA) and fluid antenna \cite{wong2020fluid} (FA), have been recently recognized as a promising approach that allows antennas to move within a local region to enhance wireless communication performance.
As a novel paradigm characterized by position and shape flexibility, flexible-position antennas can be implemented with the aid of mechanical and electronic drivers\cite{ning2024movable}.
By leveraging the ability of position refinement, MAs can proactively reshape wireless channels in a more favorable condition.
Recent results have demonstrated the great potential of MA-enabled systems in received signal power improvement \cite{zhu23}, interference mitigation \cite{xiao2024multiuser}, flexible beamforming \cite{yang2025flexible,chen2023joint}, and multiplexing enhancement \cite{ma23}.


Specifically, in \cite{zhu23}, the work studied and characterized the continuous variation of the channel power gain with respect to the MA's position. Also, in \cite{ma23}, the channel is reconfigured to maximize the capacity of MA-enabled point-to-point MIMO communication systems, which demonstrated that the spatial multiplexing performance of MIMO systems can be improved by jointly designing the position of transmit and receive MAs.
Motivated by these early studies, researchers have devoted considerable efforts to exploring the further potential of MA/FA in a wider range of applications, including intelligent reflection surface (IRS) \cite{wei2024movable}, near-field sensing \cite{wang2025antenna}, mobile edge computing (MEC) \cite{zuoJin24}, among others  \cite{chen2025energy}. 
Moreover, the performance advantages of MAs over FPAs in multiuser communication systems have also been validated. For instance, in \cite{zhu2023movable}, the MA-enhanced multi-access channel (MAC) was investigated, which shows that, by optimizing each user's MA position, the proposed algorithm can effectively reduce the correlation between users' channel vectors. In \cite{xiao2024multiuser}, a two-loop iterative algorithm based on particle swarm optimization (PSO) was proposed to maximize the minimum achievable rate, which enhances the user fairness performance in the uplink transmission. The work \cite{cheng2023sum} developed an efficient fractional programming (FP)-based position optimization algorithm to maximize the sum-rate for multi-user systems. Instead of optimizing the MA positions in a continuous space as above, \cite{wu2023movable,meizhang2024,wu2024globally} considered MA-aided systems under the constraint that MA elements are confined to discrete movements. Even though the position discretization leads to a performance loss, the works \cite{wu2023movable,meizhang2024,wu2024globally} showed that the proposed MA-enabled systems still achieve significant performance gains over the conventional FPAs.

Despite the above results, MA position optimization, which is the key to unlock the full potential of MA-enabled systems, relies on the complete knowledge of the channel state information (CSI) between
the transmitter (Tx) and the receiver (Rx) over the entire
movable regions. Traversing positions over the entire
movable regions to measure the CSI, however, leads to an excessive amount of training overhead, which is infeasible in practice. 
To address this difficulty, instead of estimating the channel response over the entire movable regions, some works (e.g., \cite{shao20256d,hu2024secure,zhu2024performance}) assume a multi-path far-field response channel model. Thus, to recover the CSI over the entire movable regions, one only needs to estimate a finite number of path parameters (e.g. the complex channel gain, the angle of departure (AoD) and angle of arrival (AoA) associated with each path component of the Tx-Rx channel. In \cite{ma2023compressed}, a compressed sensing (CS)-based method was proposed to sequentially estimate the  AoDs ,  AoAs and complex channel gains of each path component by exploiting the sparse scattering structure of the wireless channel. 
To further enhance the estimation accuracy, the work \cite{xiao2024channel} developed an improved CS-based method that jointly estimates the AoAs, AoDs, and path gains.
However, this approach suffers from computational inefficiency and is prone to grid mismatch errors, thereby limiting its channel estimation accuracy.
To address this issue, some gridless methods were proposed for MA-enabled systems to achieve a high estimation accuracy with low pilot training overhead, such as the tensor decomposition-based channel estimation/reconstruction method \cite{zhang2024channel}, the sparse channel reconstruction \cite{xu2023channel}, and the neural network-based alternating refinement algorithms\cite{jang2025new}.



The above CSI-estimation-based position optimization approach, albeit effective, suffers several major drawbacks in practical systems. First, since it requires to estimate the parameters associated with each path component to recover the global CSI over the entire movable regions, it may involve a prohibitively high amount of training overhead for rich scattering scenarios consisting of a
large number of multi-path components (MPCs). Second, the performance of this position optimization approach relies on the accuracy of the estimated parameters. As a result, it may suffer a considerable performance loss if the signal-to-noise ratio is low or the number of channel measurements is insufficient to yield an accurate estimation of the path parameters. Lastly, this two-stage approach which first estimates the path parameters and then optimizes the MA positions is computationally inefficient since both stages may incur a high computational complexity.  


In this paper, we propose a derivative-free-based approach to circumvent the above drawbacks of existing CSI-estimation-based position optimization methods.
The basic idea of the derivative-free approach is to treat position optimization as a closed-box optimization
problem and calculate the gradient of the unknown objective function using zeroth-order (ZO) gradient approximation techniques. Specifically, the proposed method adaptively refines its next movement based on previous measurements such that it eventually converges to an optimum or stationary solution, without the need of
explicitly estimating the path parameters. Thus, the proposed derivative-free approach is able to achieve a higher sample efficiency than the CSI-estimation-based position optimization approach, particularly for challenging scenarios where the number of MPCs is large.     

The current work is an extension of our prior work \cite{zeng2024csi} which considers position optimization for single-input-single-output (SISO) systems. This work extends the SISO systems to multi-user MISO systems in which the BS, equipped with multiple MAs, communicates with multiple single-FPA users. The objective is to optimize multiple MAs' positions to maximize the uplink transmission performance. Note that such an extension is nontrivial as MISO systems involve position optimization of multiple MAs, which is more complex than single MA position optimization. In addition, for multi-user scenarios, position optimization is coupled with the combining matrix design, which makes the problem much more challenging than that in SISO systems.

The rest of this paper is organized as follows.
Section \ref{section2} introduces the system model and the problem formulation of the MA-assisted MISO system. In Section \ref{section3}, we propose a derivative-free position optimization approach for single-user scenarios. The proposed method is extended to multi-user scenarios in Section \ref{section4}. Simulations results are provided in Section \ref{section5}, followed by concluding remarks in Section \ref{section6}.


\begin{figure}[t!]
	\setlength{\abovedisplayskip}{1pt}
	\setlength{\belowdisplayskip}{1pt}
	\centering
	\includegraphics[width=9cm]{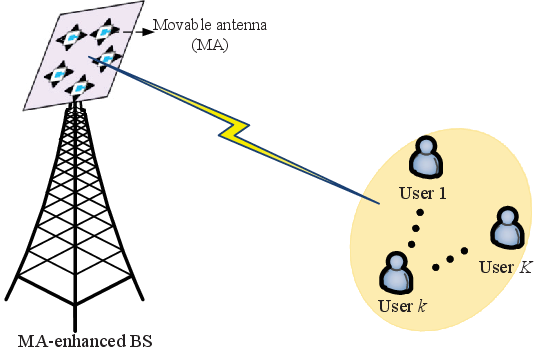}
	\caption{Illustration of the considered MA-enabled communication system}
	\label{system}
\end{figure}

\begin{figure*}[t!]
	\setlength{\abovedisplayskip}{1pt}
	\setlength{\belowdisplayskip}{1pt}
	\centering
	\includegraphics[width=16cm]{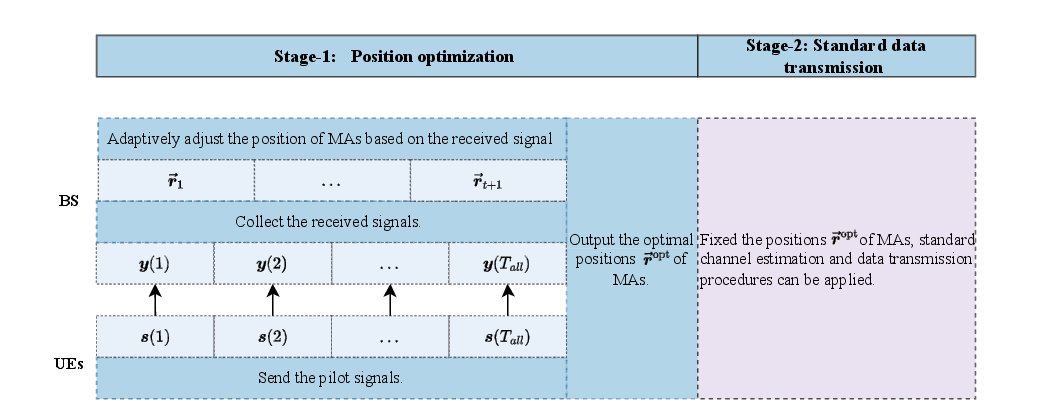}
	\caption{General framework for the proposed method}
	\label{diagram}
\end{figure*}

\section{System Model and Problem Formulation}\label{section2}
\subsection{System Model}
We consider an MA-enabled multi-user communication system, where a BS equipped with $M$ MAs serves $K$ users, with each user equipped with a single FPA (see Fig. \ref{system}). Note that we assume MAs are only deployed at the BS, as it has a less stringent power/cost constraint and can  employ a high-precision linear driver to enable continuous and precise movement of the antenna. Specifically, each MA at the BS is connected with a radio frequency (RF) chain via a flexible cable, and the position of each MA can be flexibly adjusted within a given two-dimensional (2D) region $\mathcal{R}$. Without loss of generality, we assume that $\mathcal{R}$ is a square area of size $A\times A$. A 2D local coordinate system is established to describe the position of the $m$th MA at the BS, which is denoted as $\bm{r}_m=[x_m\phantom{0}y_m]^T$. 

Consider an uplink transmission scenario where $K$ users simultaneously transmit their signals to the BS. Let $\bm{s}\in \mathbb{C}^K$ denote the transmitted signals from $K$ users, where $\mathbb{E}[\bm{s}\bm{s}^H]=\bm{I}_K$. The received signal at the BS can be expressed as
\begin{align}
\bm{y}(t)=\bm{H}\bm{s}(t)+\bm{n}(t) \in \mathbb{C}^{M}, \label{eqn1}
\end{align}
where $\bm{H}\triangleq[\bm{h}_1\phantom{0}\bm{h}_2\phantom{0}\cdots\phantom{0}\bm{h}_K]\in \mathbb{C}^{M\times K}$ is the channel matrix from $K$ users to the BS, $\bm{n}(t)\sim \mathcal{CN}(0,\sigma^2\bm{I}_M)$ is the additive white Gaussian noise. 

The objective is to optimize MAs' positions to maximize the uplink transmission performance. Note that the channel between the BS and the $k$th user, $\boldsymbol{h}_k$, is a function of MAs' positions. Define 
\begin{align}
\bm{\vec{r}}\triangleq[\bm{r}_1^T\phantom{0}\bm{r}_2^T\phantom{0}\cdots\phantom{0}\bm{r}_M^T]^T.  
\end{align}
We use $\boldsymbol{h}_k(\bm{\vec{r}})$ to indicate the dependence of the channel on the MAs' positions. To optimize MAs' positions, we consider a transmission protocol as illustrated in Fig. \ref{diagram}. Specifically, the transmission protocol consists of a position optimization stage, and a standard channel estimation/data transmission stage. In the position optimization stage, users send the pilot, e.g., $\bm{s}(t),\forall t=1,\ldots,T_{all}$, to the BS, which then utilizes the received signals $\bm{y}(t)$ to adaptively adjust the position of MAs. Here, $T_{all}$ denotes the length of the transmit pilots. The goal of this paper is to develop a sample-efficient approach to determine the optimal positions of  MAs in this stage. Once the MAs' positions are optimized and fixed, standard channel estimation and data transmission procedures can be applied.


\subsection{Channel Model}
Let $L_k$ denote the total number of signal paths from the $k$th user to the BS. 
Also, let ${h}_k(\boldsymbol{r}_o)$ denote the channel response between the $k$th user and the MA placed at the reference position $\boldsymbol{r}_o=[0\phantom{0}0]^T$. The channel response ${h}_k(\boldsymbol{r}_o)$ is a superposition of multi-path components (MPCs) and can be written as
\begin{align}
{h}_k(\boldsymbol{r}_o)=\sum_{l=1}^{L_k}b_{l,k}=\boldsymbol{1}_{L_k}^H\boldsymbol{\mu}_k,
\end{align}
where $b_{l,k}$ represents the complex path gain associated with the $l$th path of the $k$th user's channel, $\boldsymbol{\mu}_k\triangleq [b_{1,k}\phantom{0}b_{2,k}\phantom{0}\cdots\phantom{0} b_{L_k,k}]^T$ denotes the path response vector, and
$\boldsymbol{1}_{L_k}\in \mathbb{R}^{L_k}$ denotes a vector with all
entries equal to one.

Since the size of the movable region is much smaller than the propagation distance between the transmitter (Tx) and the receiver (Rx), the far-field condition is generally satisfied. As each antenna moves from the reference point $\boldsymbol{r}_o$ to position $\bm{r}_m=[x_m\phantom{0}y_m]^T$, the AoAs and the complex gain of each path component remain approximately constant over the entire movable region, whereas only the signal propagation distance of each path changes, which in turn results in a phase variation. Let $\theta_{l,k}$ and $\phi_{l,k}$ respectively denote the elevation AoA and the azimuth AoA associated with the $l$th path of the $k$th user's channel. 
It can be readily verified that, as the antenna moves from $\boldsymbol{r}_o$ to $\bm{r}_m$, the signal propagation distance of each path changes by 
\begin{align}
d(\bm{r}_m;\theta_{l,k},\phi_{l,k}) =x_m \cos\theta_{l,k}\sin\phi_{l,k}+y_m\sin\theta_{l,k}.
\end{align}
This change in distance, in turn, results in a phase variation given by $\frac{2\pi}{\lambda}d(\bm{r}_m;\theta_{l,k},\phi_{l,k})$, where $\lambda$ is the wavelength of the signal. As a result, the channel response between the $k$th user and the MA positioned at $\bm{r}_m$ can be given as
\begin{align}
h_k(\boldsymbol{r}_{m})=(\boldsymbol{p}_k(\boldsymbol{r}_m))^H\boldsymbol{\mu}_k\label{channel},
\end{align}
where $\boldsymbol{p}_k(\boldsymbol{r}_m)\in\mathbb{C}^{L_k}$ is the $k$th user's phase variation vector with its $l$th entry corresponding to the phase variation of the $l$th signal path as the antenna moves to position $\boldsymbol{r}_m$:
\begin{align}
\boldsymbol{p}_k(\boldsymbol{r}_m)
\triangleq\left[e^{j\frac{2\pi}{\lambda}d(\boldsymbol{r}_m;\theta_{1,k},\phi_{1,k})}\phantom{0}  \cdots\phantom{0}  e^{j\frac{2\pi}{\lambda}d(\boldsymbol{r}_m;\theta_{L_k,k},\phi_{L_k,k})}	\right]^T. \label{eqn3}
\end{align}

Accordingly, the channel $\boldsymbol{h}_k(\bm{\vec{r}})$ between the $k$th user and the BS can be expressed as 
\begin{align}
		\boldsymbol{h}_k(\bm{\vec{r}})
		&\triangleq \left[h_k(\boldsymbol{r}_{1})\phantom{0}  \cdots\phantom{0}h_k(\boldsymbol{r}_{M}) \right]^T \nonumber\\
		&=\left(\boldsymbol{P}_k(\bm{\vec{r}})\right)^H\boldsymbol{\mu}_k,
\end{align}
where 
\begin{align}
	\boldsymbol{P}_k(\boldsymbol{\vec{r}})\triangleq\left[\boldsymbol{p}_k(\boldsymbol{r}_1)\phantom{0}\cdots\phantom{0}\boldsymbol{p}_k(\boldsymbol{r}_M)\right]\in \mathbb{C}^{L_k\times M}
\end{align}
is the $k$th user's phase variation matrix. 

So far we have established the mathematical relationship between the BS-user channel $\boldsymbol{h}_k$ and MAs' positions $\bm{\vec{r}}$. 
From our above analysis, it is evident that the BS-user channel $\boldsymbol{h}_k$ is not only a function of MAs' positions, but also a function of the path-related parameters including the elevation and azimuth AoAs
$\{\theta_{l,k},\phi_{l,k}\}$ and the complex attenuation coefficients $\{b_{l,k}\}$. To illustrate this explicit dependence of $\boldsymbol{h}_k$ on both the MA's positions as well as the channel path's parameters, we can write $\boldsymbol{h}_k$ as $\boldsymbol{h}_k(\bm{\vec{r}};\bm{\omega}_k)$, where
$\bm{\omega}_k\triangleq\{\theta_{l,k},\phi_{l,k},b_{l,k}\}_{l=1}^{L_k}$ denotes the set of paths' parameters associated with the $k$th user.
As will be shown later, such an explicit dependence on $\bm{\omega}_k$ will help facilitate the exposition of our proposed method.


\subsection{Problem Formulation}
As the channel between each user and the BS is a function of MAs' positions, our objective is to optimize MAs' positions to maximize the uplink transmission performance\footnote{Due to the uplink-downlink channel reciprocity, the downlink performance can be optimized as well}. Here we use the sum rate as a metric to evaluate the uplink performance of MA-enabled communication systems. Upon receiving signals from the users, the BS uses a combining matrix to process the received signals, i.e.,
\begin{align}
	\bm{\hat{s}} =	 \bm{W}^H\bm{H}\bm{s}+	\bm{W} ^H\bm{n} \in \mathbb{C}^{K}, \label{eqn2}
\end{align}
where $\boldsymbol{W}=\begin{bmatrix}\boldsymbol{w}_{1}&\boldsymbol{w}_{2}&\ldots&\boldsymbol{w}_{K}\end{bmatrix}\in\mathbb{C}^{M\times K}$ denotes the combining matrix. The receive signal-to-interference-plus-noise ratio (SINR) for the $k$th user can thus be given by
\begin{align}
\gamma_k=\frac{|\boldsymbol{w}^H_{k}\boldsymbol{h}_k(\boldsymbol{\vec{r}};\bm{\omega}_k)|^2}{\sum_{i=1, i\ne k}^{K}|\boldsymbol{w}^H_{k}\boldsymbol{h}_i(\boldsymbol{\vec{r}};\bm{\omega}_k)|^2+\|\boldsymbol{w}^H_{k}\|^2_2\sigma^2}, \label{eqn4}
\end{align}
Based on (\ref{eqn4}), the achievable rate for the $k$th user is calculated as 
\begin{align}
R_k=\log_2\left(1+\gamma_k\right).
\end{align}
and the sum rate can be expressed as 
\begin{align}
	R=\sum_{k=1}^{K}R_k  \label{eqn5}
\end{align}

In this paper, we aim to maximize the sum rate by jointly optimizing the MAs' positions and the combining matrix, subject to a minimum distance constraint placed on MAs. Specifically, the optimization problem can be formulated as 
\begin{align}
\max_{\boldsymbol{W},\boldsymbol{\vec{r}}} &\quad R, \nonumber\\ 
\text{s.t.}&\quad 	\boldsymbol{r}_m\in\mathcal{R}, \qquad 1 \leq m \leq M, \nonumber\\ 
&\quad\left\|\boldsymbol{r}_i-\boldsymbol{r}_j\right\|_2\geq d,  \qquad  1 \leq i\neq j \leq M,  \label{opt1}
\end{align}
where the first constraint ensures that each MA is located within the movable region $\mathcal{R}$,  and the second constraint specifies a minimum distance $d$ between any two MAs to mitigate coupling effects. 

From (\ref{opt1}), we see that the optimization of the MAs' positions requires the knowledge of the channel paths' parameters
$\bm{\omega}\triangleq\{\bm{\omega}_k\}_{k=1}^{K}$. Existing methods for position optimization typically follow a two-step approach: estimating the channel path parameters $\bm{\omega}$ in the first step and then optimizing the MAs' positions in the second step. To estimate the channel path parameters, one needs to measure the channels at different 
positions and collects the measured channel samples. Based on these measurements, the path parameters can be estimated via compressed sensing or least-squares techniques. Such a CSI estimation-based approach, however, suffers several drawbacks. 
First, wireless propagation environments over the sub-6G frequencies are generally rich scattering with a large number of MPCs. 
In this case, estimating their associated path parameters is a challenging task and requires an excessively large number of channel measurements.
Second, parameter estimation inevitably incurs estimation errors, which has a detrimental effect on the subsequent position optimization. 

To overcome the drawback of existing approaches, in this paper we propose a derivative-free approach
which eliminates the need of the channel path knowledge for position optimization.

\section{Position Optimization for Single User MA-Aided Systems}\label{section3}
To illustrate our basic idea, we first consider the special case with a single user in the system, i.e. $K=1$. The received signal at the BS
can be written as
\begin{align}
\boldsymbol{y}(\boldsymbol{\vec{r}})=\boldsymbol{h}(\boldsymbol{\vec{r}};\bm{\omega})s+ \boldsymbol{n},
\end{align} 
where $\boldsymbol{h}(\boldsymbol{\vec{r}};\bm{\omega})\in \mathbb{C}^{M} $ is the channel between the BS and the user. For the single-user scenario, the objective function of (\ref{opt1}) can be simplified as 
\begin{align}
R=\log_2\left(1+\frac{ |\boldsymbol{w}^H\boldsymbol{h}(\boldsymbol{\vec{r}};\bm{\omega})|^2}{\sigma^2\|\boldsymbol{w}\|^2_2}\right),
\end{align}
Since logarithmic functions are monotonically increasing functions, maximizing $R$ is equivalent to maximizing the signal-to-noise ratio (SNR), which is defined as 
\begin{align}
\gamma\triangleq\frac{|\boldsymbol{w}^H\boldsymbol{h}(\boldsymbol{\vec{r}};\bm{\omega})|^2}{\sigma^2\|\boldsymbol{w}\|^2_2},
\end{align} 
According to the Cauchy–Schwarz inequality, we have
\begin{align}
\gamma=\frac{ |\boldsymbol{w}^H\boldsymbol{h}(\boldsymbol{\vec{r}};\bm{\omega})|_2^2}{\sigma^2\|\boldsymbol{w}\|^2_2} &\leq\frac{ \|\boldsymbol{w}\|_2^2\|\boldsymbol{h}(\boldsymbol{\vec{r}};\bm{\omega})\|_2^2}{\|\boldsymbol{w}\|^2_2\sigma^2}\nonumber\\
&=\frac{\|\boldsymbol{h}(\boldsymbol{\vec{r}};\bm{\omega})\|_2^2}{\sigma^2},
\end{align} 
The equality in the Cauchy–Schwarz inequality holds if and only if $\boldsymbol{w}=c\boldsymbol{h}(\boldsymbol{\vec{r}};\bm{\omega})$, where $c$ is a nonzero constant. Thus, problem (\ref{opt1}) can be reduced to a position optimization problem as 
\begin{align}
(\mathrm{P1}) \quad 	\max_{\boldsymbol{\vec{r}}}& \quad f(\boldsymbol{\vec{r}};\bm{\omega})=\|\boldsymbol{h}(\boldsymbol{\vec{r}};\bm{\omega})\|_2^2
	\nonumber	\\
 \text{s.t.} 	
&\quad \boldsymbol{r}_m\in\mathcal{R} , \qquad 1 \leq m \leq M,\nonumber\\
&\quad \left\|\boldsymbol{r}_i-\boldsymbol{r}_j\right\|_2\geq d, \qquad 1 \leq  i\neq j \leq M,	\label{P1}	
\end{align}

Note that if the channel path parameters $\bm{\omega}$ are known \emph{a priori}, then this position optimization problem becomes a conventional optimization problem that can be solved via standard optimization methods. Nevertheless, as discussed earlier, estimating the channel path parameters $\bm{\omega}$ is challenging in practice and could entail a substantial amount of training overhead, particularly for rich scattering environments. 

To address this challenge, we propose to optimize MAs' positions without explicitly estimating the parameters $\bm{\omega}$. The core idea is to treat position optimization as a derivative-free optimization (also known as black-box optimization\cite{chenLiu19}). Conventional derivative-based algorithms use derivative information of the objective function $f$ to find a good search direction. However, they have problems when the objective function $f$ is expensive to evaluate, or the function $f$ is unknown (or has unknown parameters) as in our case. Derivative-free optimization methods address these challenges using only function values of $f$ to find optimal points. 

%

Although there are different approaches (such as Bayesian optimization \cite{shahriariBobak15}, derivative-free trust region methods \cite{larsonJeffrey19}, and genetic algorithms \cite{berahasCao22}) to address the derivative-free optimization problem, among them,
the zeroth-order (ZO) optimization method \cite{chenLiu19,liuChen20} has gained an increasing attention due to its unique advantages.  First, ZO methods are straightforward to implement as they are based
on commonly used gradient-based algorithms; second, ZO
methods can achieve convergence rates comparable to first-order algorithms.

\subsection{Proposed ZO Optimization Method}
The basic idea of ZO optimization is to approximate the full gradients or stochastic gradients through function values.
Specifically, the directional derivative approximation of $f(\boldsymbol{\vec{r}})$\footnote{For simplicity, we drop the explicit dependence of $f$ on $\bm{\omega}$.} can be expressed as 
\begin{align}
\hat{\nabla}
	f(\boldsymbol{\vec{r}})=(2M/\mu)[f(\boldsymbol{\vec{r}}+\mu\boldsymbol{u})-f(\boldsymbol{\vec{r}})]
	\boldsymbol{u}, 
\end{align}
where $\boldsymbol{u}$ is a random vector drawn from the sphere of
a unit ball, and $\mu$ is a small step size. In some cases,
$\boldsymbol{u}$ can be also randomly chosen as a standard unit
vector $\boldsymbol{e}_i$ with 1 at its $i$th coordinate and zeros
elsewhere. Recall that $f(\boldsymbol{\vec{r}})=\|\boldsymbol{h}(\boldsymbol{\vec{r}};\bm{\omega})\|_2^2$. Although $\boldsymbol{h}(\boldsymbol{\vec{r}};\bm{\omega})$
is not directly available, by letting $s=1$ during the position optimization stage, we can use $\boldsymbol{y}(\boldsymbol{\vec{r}})$ to replace $\boldsymbol{h}(\boldsymbol{\vec{r}};\bm{\omega})$ in calculating the approximate derivative of $f(\boldsymbol{\vec{r}})$, where $\boldsymbol{y}(\boldsymbol{\vec{r}})$ denotes the received pilot signal at the BS when MAs are positioned at $\boldsymbol{\vec{r}}$. This leads to:
\begin{align}
\tilde{\boldsymbol{g}} =& \hat{\nabla}  f(\boldsymbol{\vec{r}})
= \frac {2M(f(\boldsymbol{\vec{r}}+\mu\boldsymbol{u})-f(\boldsymbol{\vec{r}}))}{ \mu}\boldsymbol{u} \nonumber\\
	\overset{(a)}{\approx}& \frac {2M\left(\left\|\bm{y}(\boldsymbol{\vec{r}}+\mu \boldsymbol{u})\right\|_2^2 -
			\left\|\bm{y}(\boldsymbol{\vec{r}})\right\|_2^2\right)}{\mu
		}\boldsymbol{u}, \label{eqn18}
\end{align}
Note that the gradient estimate $\tilde{\boldsymbol{g}}$ becomes more accurate when a smaller $\mu$ is adopted. On the other hand, if $\mu$ is set too small, the function
difference could be dominated by the noise and thus yields a poor
gradient estimate. Thus, a proper choice of the parameter $\mu$ is
important for the convergence of ZO.  

Most ZO optimization methods mimic their first-order counterparts
and involve three steps, namely, gradient estimation, descent
direction computation, and point updating. The gradient estimation
can be performed using (\ref{eqn18}). For different ZO methods,
their major difference lies in the strategies used to form the
descent direction. Note that the estimated gradient (\ref{eqn18})
is stochastic in nature and may suffer from a large estimation
variance, which causes poor convergence performance. To address
this issue, different descent direction update schemes have been
proposed. Among them, the ZO-AdaMM \cite{chenLiu19} has been
proven to be an effective method that is robust against gradient
estimation errors and achieves a superior convergence speed. Due
to its simplicity and superior performance, here we adopt ZO-AdaMM
to solve our position optimization problem (\ref{P1}).

For clarity, we summarize the ZO-AdaMM in Algorithm \ref{zoadamm1}. In the algorithm, the zero-order gradient is estimated via (\ref{eqn18}). Then, the descent direction $\boldsymbol{m}_t$ is computed by an exponential moving average of the past gradients. In addition to the descent direction, ZO-AdaMM calculates a second
moment vector $\boldsymbol{v}_t$ that
dynamically adjusts the step size. Finally, after the descent direction and the second moment vector
are calculated, the MA position can be updated as $\boldsymbol{\vec{r}}_{t+1}= \boldsymbol{\vec{r}}_{t}+\alpha\hat{\boldsymbol{V}}_{t}^{-\frac{1}{2}}{\hat{\boldsymbol{m}}_{t}}$.

\begin{algorithm}[htb]
	\renewcommand{\algorithmicrequire}{\textbf{Input:}}
	\renewcommand{\algorithmicensure}{\textbf{Output:}}
	\caption{ZO-AdaMM }
	\label{zoadamm1}
	\begin{algorithmic}[1]
		\REQUIRE  step size $\alpha$, hyper-parameters
		$\beta_{1}$, $\beta_2\in \left(0,1\right] $, and $\boldsymbol{m}_0$, $\boldsymbol{v}_0$ ~~\\
		\ENSURE $\boldsymbol{\vec{r}}^{*}$\\
		\STATE { Initialize the MA position $\boldsymbol{r}_{0}$.}
		\WHILE {not converge}
		\STATE {Randomly pick coordinate or vector $\boldsymbol{u}$. }
		\STATE {$t \gets t +1 $ }
		\STATE {Estimate $\tilde{\boldsymbol{g}}_t= \hat{\nabla}f(\boldsymbol{x})$  according to (\ref{eqn18})}
		
		\STATE {let     $\boldsymbol{m}_{t}=\beta_1 \boldsymbol{m}_{t-1}
			+(1-\beta_1)\tilde{\boldsymbol{g}}_t$} \STATE {
			$\boldsymbol{v}_{t}=\beta_2 \boldsymbol{v}_{t-1}
			+(1-\beta_2)\tilde{\boldsymbol{g}}_t\circ\tilde{\boldsymbol{g}}_t$\\}
		\STATE {    $\hat{\boldsymbol{m}}_{t}=\frac{\boldsymbol{m}_{t}}{1-\beta_1^t}$}
		\STATE {
			$\hat{\boldsymbol{v}}_{t}=\frac{\boldsymbol{v}_{t}}{1-\beta_2^t}$,
			and
			$\hat{\boldsymbol{V}}_{t}=\text{diag}(\hat{\boldsymbol{v}}_{t})$ }
		\STATE {Update $\boldsymbol{\vec{r}}_{t+1}\gets  \boldsymbol{\vec{r}}_{t}+
			\alpha\hat{\boldsymbol{V}}_{t}^{-\frac{1}{2}}{\hat{\boldsymbol{m}}_{t}} $
		}
		\ENDWHILE
		\STATE{Obtain the optimal solution $\boldsymbol{\vec{r}}^{*}=\boldsymbol{\vec{r}}_{t+1} $ }
		\RETURN{$\boldsymbol{\vec{r}}^{*}$}
	\end{algorithmic}
\end{algorithm}

Note that ZO-AdaMM is designed to solve unconstrained optimization problems. Nevertheless, our problem (P1) is a constrained problem which is subject to a feasible region constraint as well as a minimum distance constraint. To address the feasible region constraint, we incorporate a projection operation during the iterative process. Specifically, if an MA moves out of the boundary of the feasible region, we project its position component to the corresponding boundary point, i.e.
\begin{equation}
	 \left[\boldsymbol{\mathcal{B}}\left(\boldsymbol{\vec{r}}\right)\right]_i=\left\{
	\begin{aligned}
		-\frac{A}{2} \quad  &\text{if $\left[\boldsymbol{\vec{r}}\right]_i <-\frac{A}{2}$},\\
		\frac{A}{2}\quad     & \text{if $\left[\boldsymbol{\vec{r}}\right]_i >\frac{A}{2}$}, \\
		\left[\boldsymbol{\vec{r}}\right]_i\quad  & \text{otherwise}. \\
	\end{aligned} \label{boundary}
	\right.
\end{equation}
This projection operation ensures that each MA is always located within the feasible region during the iteration process. 

On the other hand, for the minimum distance constraint, we temporarily relax this constraint during the iterative process. After the algorithm converges, we refine the obtained MA positions to satisfy this constraint, as detailed next. Let $\mathcal{L}_1$ and $\mathcal{L}_2$ denote the sets of MA indices that violate and satisfy the minimum distance constraint, respectively. If the set $\mathcal{L}_1$ is empty, then we just keep the optimized positions $\boldsymbol{ \vec{r}}^*$ unchanged. If the set $\mathcal{L}_1$ is nonempty, we first construct a uniform grid that covers the entire movable region $\mathcal{R}$. The distance between neighboring grid points is set to $d$, i.e. the minimum allowable distance between antennas. Specifically, the grid is given by 
\begin{equation}
	\mathcal{G} = \left\{ \bm{g}_{k,l}\, \middle| \, \bm{g}_{k,l}=\bm{o} + k d \hat{\bm{x}} + l d \hat{\bm{y}},    \, 
	k,l \in \mathbb{Z}, 		 \|\bm{g}_{k,l}\| \in \mathcal{R}
	\right\}.\label{grid}
\end{equation}
where the origin $\bm{o}=[x_o,y_o]$  is randomly chosen from the movable region and the basis vectors $\hat{\bm{x}}$ and $\hat{\bm{y}}$ are standard unit vectors. 
For those antennas in the set $\mathcal{L}_2$, their positions remain unaltered. We only need to project the positions of antennas in the set $\mathcal{L}_1$ to different grid points. This position refinement is proceeded in a sequential manner, i.e., we first project the position of the first antenna in $\mathcal{L}_1$ to a certain grid point, and then the second antenna in $\mathcal{L}_1$ to another grid point, and so on and so forth. During this process, to avoid violating the minimum distance constraint, the position of each antenna in $\mathcal{L}_1$ is not allowed to be projected to those occupied grid points and to those grid points that are next to the antennas in $\mathcal{L}_2$. 

Specifically, define ${\mathcal{G}}_{\text{avail}}$ as the set of available grid points, which is dynamically updated by removing  grid points that have already been occupied, i.e., $\mathcal{G}_{\text{avail}}= \mathcal{G} \setminus \mathcal{U}$, where $\mathcal{U}$ is defined as the set of occupied grid points. Additionally, those grid points that are closest to the MAs in the set $\mathcal{L}_2$ are marked as occupied. Then, for each antenna in $\mathcal{L}_1$, say the $p$th antenna, we assign this antenna to the closest available grid point, i.e.
\begin{align}
	\bm{g}_p = \arg\min_{\bm{g} \in \mathcal{G}_{\text{avail}}} \|\bm{g} - \bm{r}_p^*\|.
\end{align}
Next, we update the set of available grid points by removing $\bm{g}_p$ and then proceed to process the $(p+1)$th antenna in $\mathcal{L}_1$. This process is continued until all antennas in $\mathcal{L}_1$ are projected to different grid points. After this process is completed, we can ensure that new positions of all MAs, including antennas in $\mathcal{L}_1$ and $\mathcal{L}_2$, will satisfy the minimum distance constraint.


For clarity, the algorithm is summarized in Algorithm 2. 

\begin{algorithm}[htb]
	\caption{Modified ZO-AdaMM for Position Optimization}
	\renewcommand{\algorithmicrequire}{\textbf{Input:}}
	\renewcommand{\algorithmicensure}{\textbf{Output:}}
	\label{summar}
	\begin{algorithmic}[1]
		\REQUIRE  step size $\alpha$, hyper-parameters
		$\beta_{1}$, $\beta_2\in \left(0,1\right] $, minimum distance $d$ and $\boldsymbol{m}_0$, $\boldsymbol{v}_0$ ~~.\\
		\ENSURE $\boldsymbol{\vec{r}}^{{\text{opt}}}$.\\
		\STATE { Randomly generate a set of candidate positions $\mathcal{R}^{ini}=\{\boldsymbol{\vec{r}}_i^{ini}\}_{i=1}^{P_i}$.}
		\STATE {Evaluate the performance of each candidate position via the received signal power.}
		\STATE {Select the position that achieves the highest received signal power as the initial point.}
		\WHILE {not converge}
		\STATE {Randomly pick a coordinate or a vector $\boldsymbol{u}$. }
		\STATE {$t \gets t +1 $. }
		\STATE {Estimate $\tilde{\boldsymbol{g}}_t= \hat{\nabla}f(\boldsymbol{x})$  according to (\ref{eqn18}).}
		\STATE {let     $\boldsymbol{m}_{t}=\beta_1 \boldsymbol{m}_{t-1}
			+(1-\beta_1)\tilde{\boldsymbol{g}}_t$} \STATE {
			$\boldsymbol{v}_{t}=\beta_2 \boldsymbol{v}_{t-1}
			+(1-\beta_2)\tilde{\boldsymbol{g}}_t\circ\tilde{\boldsymbol{g}}_t$.}
		\STATE {    $\hat{\boldsymbol{m}}_{t}=\frac{\boldsymbol{m}_{t}}{1-\beta_1^t}$}
		\STATE {
			$\hat{\boldsymbol{v}}_{t}=\frac{\boldsymbol{v}_{t}}{1-\beta_2^t}$,
			and
			$\hat{\boldsymbol{V}}_{t}=\text{diag}(\hat{\boldsymbol{v}}_{t})$. }
		\STATE {Update $\boldsymbol{\vec{r}}_{t+1}\gets 	\boldsymbol{\mathcal{B}}\left( \boldsymbol{\vec{r}}_{t}+ 	\alpha\hat{\boldsymbol{V}}_{t}^{-\frac{1}{2}}{\hat{\boldsymbol{m}}_{t}}\right)$.
		}
		\ENDWHILE
		\STATE{Obtain the optimal solution $\boldsymbol{\vec{r}}^{\text{opt}}=\boldsymbol{\vec{r}}_{t+1} $.}
			\STATE Define $\mathcal{L}_1 \gets \{p \mid \exists q, \|\bm{r}_p^{\text{opt}} - \bm{r}_q^{\text{opt}}\| < d \}$ and $\mathcal{L}_2=\{1,\ldots,M\}-\mathcal{L}_1$.
		\IF{$\mathcal{L}_1 = \emptyset$}
		\RETURN{$\boldsymbol{\vec{r}}^{opt}$}.
		\ELSE
		\STATE Construct a dynamic grid $\mathcal{G}$ via (\ref{grid}).
		
		\STATE Initialize the set of occupied grid points as $\mathcal{U} \gets \{\bm{g} \in \mathcal{G} \mid \|\bm{g} - \bm{r}_i^{\text{opt}}\| < d, \forall i \in \mathcal{L}_2\}$.

		\FOR{each $p \in \mathcal{L}_1$} 
		\STATE Compute the set of available grid points: $\mathcal{G}_{\text{avail}} \gets \mathcal{G} \setminus \mathcal{U}$.
		\STATE Find the nearest available grid point: $\bm{g}_p = \arg\min_{\bm{g} \in \mathcal{G}_{\text{avail}}} \|\bm{g} - \bm{r}_p^*\|$.
		\STATE Update $\bm{r}_p^{\text{opt}} \gets \bm{g}_p$.
		\STATE Update $\mathcal{U} \gets \mathcal{U} \cup \{\bm{g}_p\}$.
		\ENDFOR
		\RETURN{$\boldsymbol{\vec{r}}^{opt}$.}
		\ENDIF
	
	\end{algorithmic}	
\end{algorithm}

\section{Position Optimization for Multi-User Systems}\label{section4}
In this section, we extend our proposed derivative-free approach to multi-user systems. Recall that for multi-user systems, the position optimization can be formulated
as
\begin{align}
(\mathrm{P2}) \quad \max_{\boldsymbol{W},\boldsymbol{\vec{r}}} &\quad R=\sum_{k=1}^{K} R_k \nonumber\\ 
\text{s.t.}&\quad 	\boldsymbol{r}_m\in\mathcal{R}\qquad 1 \leq m \leq M, \nonumber\\ 
&\quad\left\|\boldsymbol{r}_i-\boldsymbol{r}_j\right\|_2\geq d \qquad  1 \leq i\neq j \leq M,  \label{P2}
\end{align}
where $R_k=\log_{2}(1+\gamma_k)$ is the achievable rate for the $k$th user with $\gamma_k$ defined in (\ref{eqn4}). Jointly optimizing the combining matrix $\boldsymbol{W}$ and
the MAs' positions $\boldsymbol{\vec{r}}$ is a challenging task, particularly when the knowledge of the path parameters $\boldsymbol{\omega}$ that appear in the objective function of (\ref{P2}) is unavailable. To circumvent this difficulty, we simplify this problem by devising $\boldsymbol{W}$ as a minimum mean square error (MMSE) estimator whose objective is to estimate the transmitted symbol vector $\boldsymbol{s}$ with a minimum distortion. Recalling (\ref{eqn2}), the MMSE estimator $\boldsymbol{W}$ is a function of the channel matrix $\boldsymbol{H}(\boldsymbol{\vec{r}};\bm{\omega})$ and can be given as
\begin{align}
\boldsymbol{W}^{\star}&=\left(\boldsymbol{H}(\boldsymbol{\vec{r}};\bm{\omega})\boldsymbol{H}(\boldsymbol{\vec{r}};\bm{\omega})^H+{\sigma^2} \boldsymbol{I}\right)^{-1}\boldsymbol{H}(\boldsymbol{\vec{r}};\bm{\omega}). 
\label{MMSE-combiner}
\end{align}     
For the MMSE estimator, its MSE $e_k$ and the achievable rate $R_k$ is related as
(see Appendix \ref{appendixA})
\begin{align}
	R_k=\log\left( \left(e_k\right)^{-1}\right),\label{Rate_MSE}
\end{align}
where $e_k\triangleq \mathbb{E}\big[\left\|(\hat{s}_k- {s_k})\right\|^{2}\big] $ is the mean squared error between the estimated signal $\hat{s}_k$ and the signal $s_k$. Thus, problem (P2) can be simplified as
\begin{align}
\min_{\boldsymbol{\vec{r}}} \quad & \sum_{k=1}^K {\log(e_k)},
		\nonumber\\
\text{s.t.}  \quad&\boldsymbol{r}_m\in\mathcal{R}, \quad 1\leq m \leq {M},
		\nonumber\\
&\left\|\boldsymbol{r}_i-\boldsymbol{r}_j\right\|_2 \geq d, \quad 1 \leq i\neq j \leq M. \label{opt2}
\end{align}  
According to Jensen's inequality, the objective function in (\ref{opt2}) is upper bounded by  
\begin{align}
\sum_{k=1}^K {\log(e_k)}=\log\left(\prod_{k=1}^{K} e_k\right) \leq K\log\left(\frac{\sum_{k=1}^K {e_k}}{K}\right).
\end{align}
As the objective function in (\ref{opt2}) is intractable, we instead minimize its upper bound, which leads to
\begin{align}
\min_{\boldsymbol{\vec{r}}} \quad & \sum_{k=1}^K e_k
		\nonumber\\
\text{s.t.}  \quad&\boldsymbol{r}_m\in\mathcal{R}, \quad 1\leq m \leq {M},
		\nonumber\\
&\left\|\boldsymbol{r}_i-\boldsymbol{r}_j\right\|_2 \geq d, \quad 1 \leq i\neq j \leq M. \label{opt3}
\end{align}  
It can be easily verified that the MSE attained by the MMSE estimator is given by (see Appendix \ref{appendixB})
\begin{align}
\sum_{k=1}^K e_k=\sigma^2
\text{tr}\left[\left(\boldsymbol{H}(\boldsymbol{\vec{r}};\bm{\omega})^{H}\boldsymbol{H}(\boldsymbol{\vec{r}};\bm{\omega})+{\sigma^2}\boldsymbol{I}_K\right)^{-1}\right]
\end{align}
Eventually we can rewrite (\ref{opt3}) as 
\begin{align}
(\text{P3}) \qquad \min_{\boldsymbol{\vec{r}}} \quad & \text{tr}\left[\left(\boldsymbol{H}(\boldsymbol{\vec{r}};\bm{\omega})^{H}\boldsymbol{H}(\boldsymbol{\vec{r}};\bm{\omega})+{\sigma^2}\boldsymbol{I}_K\right)^{-1}\right]
		\nonumber\\
\text{s.t.}  \quad&\boldsymbol{r}_m\in\mathcal{R}, \quad 1\leq m \leq {M},
		\nonumber\\
&\left\|\boldsymbol{r}_i-\boldsymbol{r}_j\right\|_2 \geq d, \quad 1 \leq i\neq j \leq M. \label{P3}
\end{align} 

So far we have obtained a more tractable optimization problem for MA position optimization for multi-user scenarios. Note that if the channel path parameters $\bm{\omega}$ 
are known \emph{a priori}, Problem (P3) can be solved via standard optimization methods. Since estimating the path parameters $\bm{\omega}$ is challenging in practice, we resort to the derivative-free optimization approach that optimizes MAs' positions without requiring the knowledge of $\bm{\omega}$.

Again, we employ the ZO optimization method to solve problem (P3). Define
\begin{align}
f(\boldsymbol{\vec{r}};\bm{\omega})\triangleq \text{tr}\left[\left(\boldsymbol{H}(\boldsymbol{\vec{r}};\bm{\omega})^{H}\boldsymbol{H}(\boldsymbol{\vec{r}};\bm{\omega})+{\sigma^2}\boldsymbol{I}_K\right)^{-1}\right]. \label{function}
\end{align}
For simplicity, we drop the explicit dependence of $f(\boldsymbol{\vec{r}};\bm{\omega})$ and $\boldsymbol{H}(\boldsymbol{\vec{r}};\bm{\omega})$ on $\bm{\omega}$, and express $f(\boldsymbol{\vec{r}};\bm{\omega})$ as $f(\boldsymbol{\vec{r}})$, and $\boldsymbol{H}(\boldsymbol{\vec{r}};\bm{\omega})$ as $\boldsymbol{H}(\boldsymbol{\vec{r}})$. Similarly, the directional derivative approximation of $f(\boldsymbol{\vec{r}})$ is calculated as
\begin{align}
 	\hat{\nabla}
 f(\boldsymbol{\vec{r}})=(2M/\mu)[f(\boldsymbol{\vec{r}}+\mu\boldsymbol{u})-f(\boldsymbol{\vec{r}})]
 	\boldsymbol{u}. \label{est_gra}
\end{align}
From the above equation, we see that we will need to obtain the channel matrix $\boldsymbol{H}(\boldsymbol{\vec{r}})$ and $\boldsymbol{H}(\boldsymbol{\vec{r}}+\mu\boldsymbol{u})$ when computing the ZO gradient at point $\boldsymbol{\vec{r}}$. Although $\boldsymbol{H}(\boldsymbol{\vec{r}})$ is unavailable, it can be readily estimated from the received pilot signals. Specifically, in the position optimization stage, users send pilot signals to the BS to assist the optimization of the MA positions. The received pilot signals when MAs are placed at position $\boldsymbol{\vec{r}}$ can be expressed as
\begin{align}
\bm{y}(t;\boldsymbol{\vec{r}})=\bm{H}(\boldsymbol{\vec{r}})\bm{s}(t)+\bm{n}(t).
\end{align} 
Collecting pilot signals over $T$ time instants, we have
\begin{align}
	\boldsymbol{Y}(\boldsymbol{\vec{r}})=\boldsymbol{H}(\boldsymbol{\vec{r}})\boldsymbol{S}+\boldsymbol{N}.
\end{align}
where 
\begin{align}
\boldsymbol{Y}&\triangleq\begin{bmatrix}\boldsymbol{y}(1;\boldsymbol{\vec{r}})\phantom{0} \boldsymbol{y}(2,\boldsymbol{\vec{r}})\phantom{0}\ldots\phantom{0} \boldsymbol{y} (T;\boldsymbol{\vec{r}})\end{bmatrix}\in\mathbb{C}^{M\times T}, \\
\boldsymbol{S}&\triangleq\begin{bmatrix}\boldsymbol{s}(1)\phantom{0}\boldsymbol{s}(2)\phantom{0}\ldots\phantom{0}\boldsymbol{s}(T)\end{bmatrix}\in\mathbb{C}^{K\times T}, \\
\boldsymbol{N}&\triangleq\begin{bmatrix}\boldsymbol{n}(1)\phantom{0}\boldsymbol{n}(2)\phantom{0}\ldots\phantom{0}\boldsymbol{n}(T)\end{bmatrix}\in\mathbb{C}^{M\times T}.
\end{align}
Consequently, we can estimate the channel matrix $\boldsymbol{H}(\boldsymbol{\vec{r}})$ using the least squares (LS) method, i.e.,
\begin{align}
\hat{\boldsymbol{H}}(\boldsymbol{\vec{r}})=\min_{\boldsymbol{H}} \quad ||\boldsymbol{Y}-\boldsymbol{H}\boldsymbol{S}||^{2}_{2}.\label{LS}
\end{align} 
Based on the estimated ZO gradient, the ZO optimization method developed in the previous section can be directly applied to solve problem P3. For clarity, the algorithm is summarized in Algorithm 3.

\begin{algorithm}[htb]
	\caption{ZO-based Algorithm for solving problem (\ref{P1}) }
		\renewcommand{\algorithmicrequire}{\textbf{Input:}}
	\renewcommand{\algorithmicensure}{\textbf{Output:}}
	\label{summar1}
	\begin{algorithmic}[1]
		\REQUIRE  step size $\alpha$, hyper-parameters
		$\beta_{1}$, $\beta_2\in \left(0,1\right] $, minimum distance $d$ and $\boldsymbol{m}_0$, $\boldsymbol{v}_0$. ~~\\
		\ENSURE $\boldsymbol{\vec{r}}^{\text{opt}}$.\\
		\STATE { Randomly generate a set of candidate positions defined as  $\mathcal{R}^{ini}=\{\boldsymbol{\vec{r}}_i^{ini}\}_{i=1}^{P_i}$ .}
		\STATE {Calculate the channel matrix at each candidate position via (\ref{LS}).}
		\STATE {Based on the estimated channel matrix, evaluate the performance for each candidate position via (\ref{function}).}
		\STATE {Select the position  with the best performance as the initial position.}
		\WHILE {not converge}
		\STATE {Randomly pick a vector $\boldsymbol{u}$. }
		\STATE {$t \gets t +1 $ }
		\STATE {Calculate the channel matrix $\hat{\boldsymbol{H}}(\boldsymbol{\vec{r}})$ via the LS estimator, and then estimate $\tilde{\boldsymbol{g}}_t= \hat{\nabla}f(\boldsymbol{x})$  according to (\ref{est_gra})}.
		\STATE {let     $\boldsymbol{m}_{t}=\beta_1 \boldsymbol{m}_{t-1}
			+(1-\beta_1)\tilde{\boldsymbol{g}}_t$.} \STATE {
			$\boldsymbol{v}_{t}=\beta_2 \boldsymbol{v}_{t-1}
			+(1-\beta_2)\tilde{\boldsymbol{g}}_t\circ\tilde{\boldsymbol{g}}_t$.}
		\STATE {    $\hat{\boldsymbol{m}}_{t}=\frac{\boldsymbol{m}_{t}}{1-\beta_1^t}$.}
		\STATE {
			$\hat{\boldsymbol{v}}_{t}=\frac{\boldsymbol{v}_{t}}{1-\beta_2^t}$,
			and
			$\hat{\boldsymbol{V}}_{t}=\text{diag}(\hat{\boldsymbol{v}}_{t})$. }
		\STATE {Update $\boldsymbol{\vec{r}}_{t+1}\gets 	\boldsymbol{\mathcal{B}}\left( \boldsymbol{\vec{r}}_{t}+ 	\alpha\hat{\boldsymbol{V}}_{t}^{-\frac{1}{2}}{\hat{\boldsymbol{m}}_{t}}\right)$.
		}
		\ENDWHILE
		\STATE{Obtain the optimal solution $\boldsymbol{\vec{r}}^{\text{opt}}=\boldsymbol{\vec{r}}_{t+1} $.}
		\STATE Define $\mathcal{L}_1 \gets \{p \mid \exists q, \|\bm{r}_p^{\text{opt}} - \bm{r}_q^{\text{opt}}\| < d \}$ and $\mathcal{L}_2=\{1,\ldots,M\}-\mathcal{L}_1$.
		\IF{$\mathcal{L}_1 = \emptyset$}
			\RETURN{$\boldsymbol{\vec{r}}^{\text{opt}}$}
		\ELSE
		\STATE Construct a grid $\mathcal{G}$ via (\ref{grid}).
		
		\STATE Initialize the set of occupied grid points as $\mathcal{U} \gets \{\bm{g} \in \mathcal{G} \mid \|\bm{g} - \bm{r}_i^{\text{opt}}\| < d, \forall i \in \mathcal{L}_2\}$.
		\FOR{each $p \in \mathcal{L}_1$} 
		\STATE Compute the set of available grid points: $\mathcal{G}_{\text{avail}} \gets \mathcal{G} \setminus \mathcal{U}$.
		\STATE Find the nearest available grid point: $\bm{g}_p = \arg\min_{\bm{g} \in \mathcal{G}_{\text{avail}}} \|\bm{g} - \bm{r}_p^{\text{opt}}\|$
		\STATE Update $\bm{r}_p^{\text{opt}} \gets \bm{g}_p$.
		\STATE Update $\mathcal{U} \gets \mathcal{U} \cup \{\bm{g}_p\}$.
		\ENDFOR
		\RETURN{$\boldsymbol{\vec{r}}^{\text{opt}}$}.
		\ENDIF	
	\end{algorithmic}	
\end{algorithm} 


\section{Simulation Results}\label{section5}
This section presents simulation results to evaluate the proposed derivative-free position optimization method, referred to as DF-PO. 
In our experiments, we consider an MA-enabled system where $M=4$ MAs are equipped at the BS and an FPA is equipped at each user. 
Users are uniformly distributed with their distances from the BS randomly generated from 20 meters to 100 meters. 
The wireless channel between the BS and each user is generated according to (\ref{channel}). The path coefficients $\{b_{l,k}\}$ are independent and identically distributed (i.i.d.) circularly symmetric complex Gaussian random variables, i.e., $b_{l,k}\sim \mathcal{CN}(0,\mu_k^2/{L})$, where $\mu_k^2=\rho d_k^{-\alpha}$ is the expected channel gain with $d_k$ denoting the distance between the BS and the $k$th user, $\rho$ denoting the path loss at the reference distance of 1m and $\alpha$ representing the path loss exponent. The elevation AoA and the azimuth AoA associated with each path are uniformly chosen over the interval $\left[-\pi/2,\pi/2\right]$. The movable region for each MA at the BS is set as a square area with size $4\lambda\times 4\lambda$, i.e.,  $\mathcal{R}=[-2\lambda,2\lambda] \times [-2\lambda, 2\lambda]$, where $\lambda$ denotes the wavelength of a signal with a frequency of $f_c=5$GHz. The power of the observation noise is set to $\sigma^2=-90\text{dBm}$.
Also, the parameters used for our proposed method are set as follows: $\beta_1=0.9$, $\beta_2=0.99$, and $d=1/4\lambda$.



To illustrate the effectiveness of our proposed algorithm, we compare the proposed algorithm with the following benchmark schemes:
 \begin{itemize} 
\item \textbf{Upper Bound (UB)}: This scheme assumes perfect knowledge of all channel path parameters and then optimize the MAs' positions via a PSO-based optimization method \cite{xiao2024multiuser}. Clearly, this method represents an upper bound on the best achievable performance by any practical method.
\item \textbf{FPA}: The BS is equipped with a fixed-position uniform planar array (denoted as FPA) with $M$ antennas spaced by $\lambda/2$. 
\item \textbf{Random Position Selection (RPS)}: Randomly generate a number of position vectors. For each randomly generated position vector $\boldsymbol{\vec{r}}$, the channel $\boldsymbol{H}$ is estimated based on the received pilot signals. Then the combining matrix can be determined via (\ref{MMSE-combiner}), and the achievable rate can be calculated according to (\ref{eqn5}) for each position vector. The position vector that attains the largest rate is selected.  
\item \textbf{Global CSI Estimation-Based Method (CSI-EB)}: This scheme first estimates the channel path parameters based on the received pilot signals. Specifically, the MAs need to collect measurements at designated or randomly selected positions and then a compressed sensing-based method \cite{ma2023compressed} is employed to estimate the associated path parameters as well as reconstruct the global CSI. After that, a PSO-based method is employed to optimize the MAs' positions. 
\end{itemize}

\subsection{Single-User Scenario}
\begin{figure}[htb]
	\setlength{\abovedisplayskip}{1pt}
	\setlength{\belowdisplayskip}{1pt}
	\centering
	\includegraphics[width=9cm]{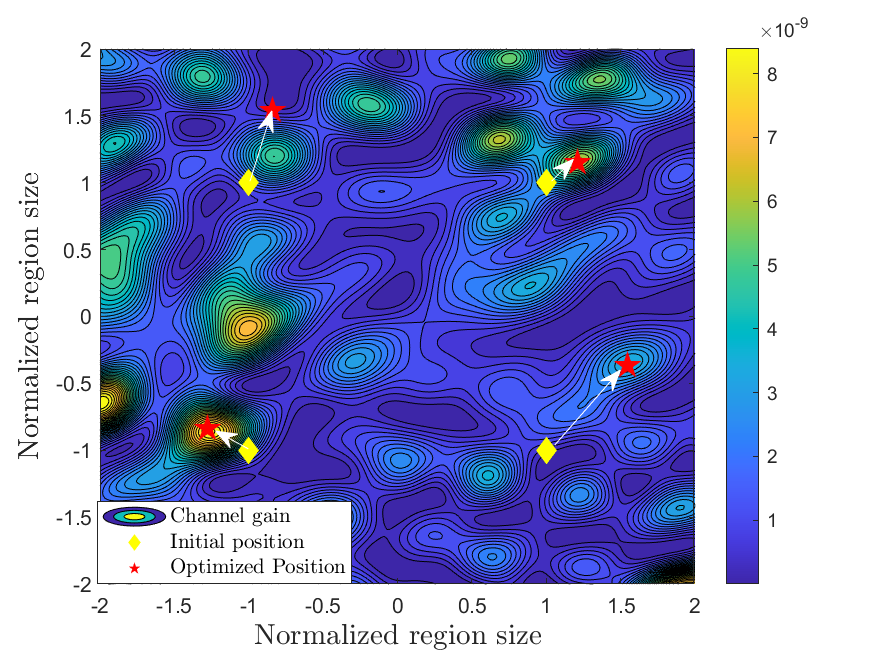}
	\caption{Variation of the receive SNR over the movable region.}
	\label{fig:1}
\end{figure}

We first consider the single-use scenario. Fig. \ref{fig:1} depicts the variation of the receive SNR over the entire movable region. The movable region is set as a 2D square area $[-2\lambda, 2\lambda]\times[-2\lambda, 2\lambda]$.  It can be seen that the movement of the MA results in the change of the receive SNR due to the small-scale fading in the spatial domain. In the figure, the positions of the $4$ MAs are randomly initialized and the initial position vector corresponds to an achievable rate of $6.3$ bps/Hz. After a few iterations, our proposed algorithm converges to positions marked by red stars as shown in Fig. \ref{fig:1}, which yields an achievable rate of $7.8$ bps/Hz. 

\begin{figure}[htb]
	\setlength{\abovedisplayskip}{1pt}
	\setlength{\belowdisplayskip}{1pt}
	\centering
	\includegraphics[width=9cm]{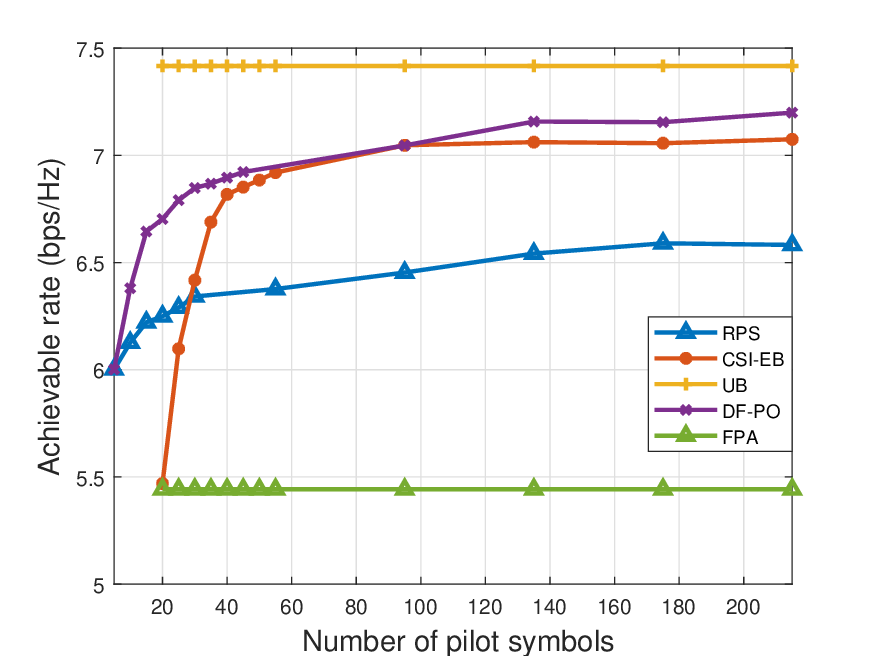}
	\caption{Achievable rates of different methods versus the number of pilot symbols.}
	\label{fig:2}
\end{figure}

\begin{figure}[htb]
	\setlength{\abovedisplayskip}{1pt}
	\setlength{\belowdisplayskip}{1pt}
	\centering
	\includegraphics[width=9cm]{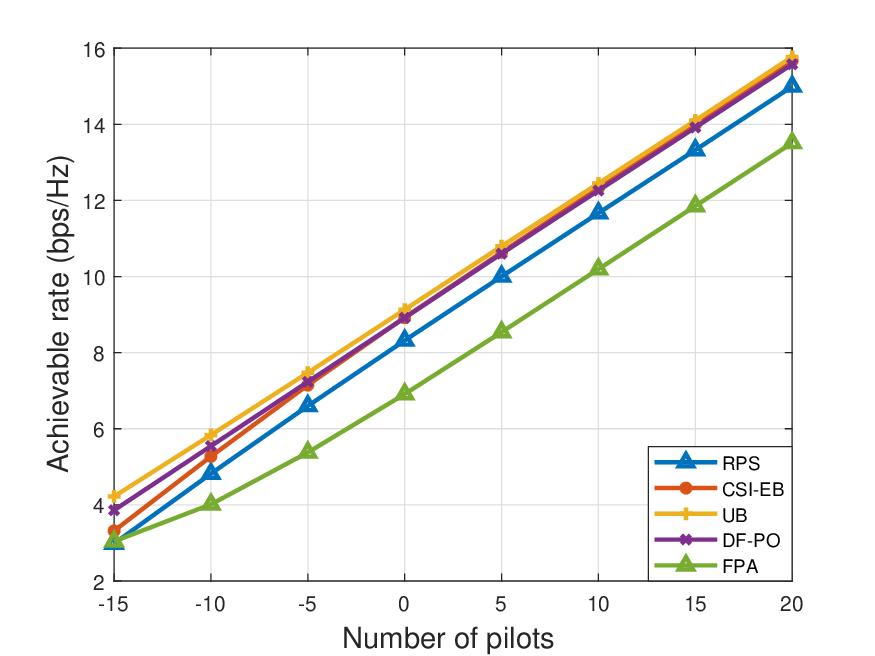}
	\caption{Achievable rates of different method versus the transmit power.}
	\label{fig:3}
\end{figure}




We now compare the sample efficiency performance of respective algorithms. For the single-user scenario, our proposed method requires a total number of pilot symbols $T_{all}=T_{ini}+T_{opt}$ for position optimization, where $T_{ini}\triangleq P_i$ pilots are used to determine a good initialization point with $P_i$ denoting the number of candidate positions and the rest $T_{opt}\triangleq 2P_z$ pilots are used to calculate the ZO gradients during the iterative process, with $2$ samples needed at each iteration for computing the current ZO gradient and $P_z$ denoting the total number of iterations performed by the algorithm. 
Fig. \ref{fig:2} plots the achievable rates versus the number of pilot symbols that are used for position optimization, where the transmit power and the number of path are set to $P_t=-5$dBm and $L=70$, respectively. We see that our proposed derivative-free approach presents a clear performance advantage over all competing algorithms. Particularly, the proposed method outperforms the CSI-EB method which first estimates the channel path parameters and then performs position optimization based on the estimated parameters. Such an improvement over the CSI-EB method can be explained as follows. First, our proposed method, without explicitly estimating the channel path parameters, can more efficiently utilize the measurements for position optimization. Such a merit is particularly advantageous when the number of pilot symbols is small, as observed from Fig. \ref{fig:2}. Second, the proposed method is robust against observation noise as it does not need to explicitly estimate the channel path parameters. Although the observation noise also leads to ZO gradient estimation errors, the ZO-AdaMM is known to be robust against gradient estimation errors. As a comparison, the CSI-EB method is more sensitive to noise since the noise is very likely to impair the parameter estimation accuracy, which in turn affects the position optimization performance.

Fig. \ref{fig:3} plots the achievable rates of respective methods versus the transmit power, where the number of pilot symbols is set to $T_{all}=100$. 
It can be observed that the CSI-EB method performs worse than the FPA in the low SNR regime (corresponding to a transmit power of $-20$dBm). This is because the channel path parameters cannot be reliably estimated when the SNR is low, which in turn affects its position optimization performance. 
As a comparison, the proposed DF-PO method presents a clear performance advantage over the FPA scheme in the low SNR regime, and this advantage is maintained for all considered transmit powers. This result, again, demonstrates the robustness and superiority of the proposed DF-PO method over the CSI-EB method.

\begin{figure}[t]
	\setlength{\abovedisplayskip}{1pt}
	\setlength{\belowdisplayskip}{1pt}
	\centering
	\includegraphics[width=9cm]{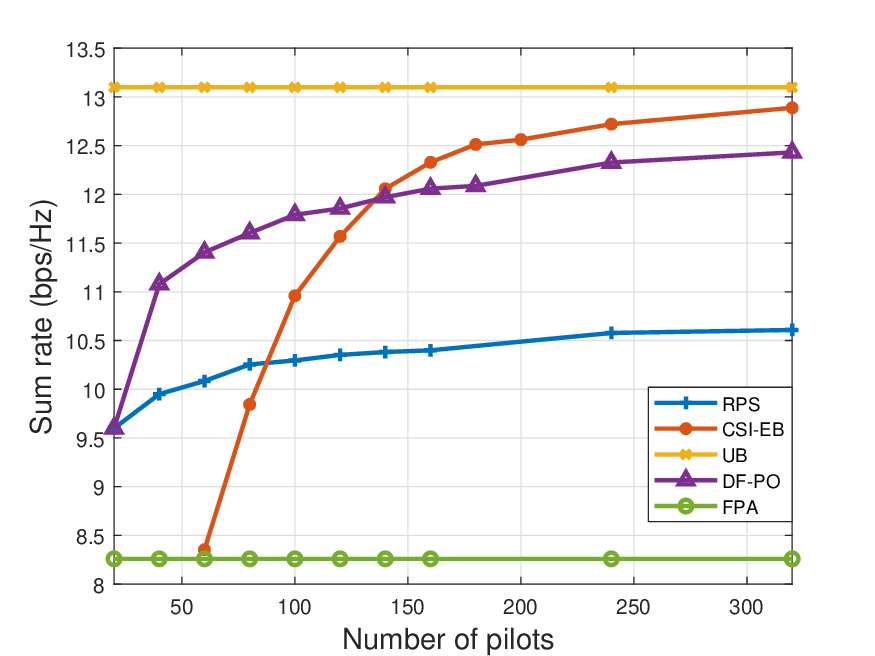}
	\caption{Sum-rates of respective methods versus the number of pilot symbols.}
	\label{fig:4}
\end{figure}

\begin{figure}[t]
	\setlength{\abovedisplayskip}{1pt}
	\setlength{\belowdisplayskip}{1pt}
	\centering
	\includegraphics[width=9cm]{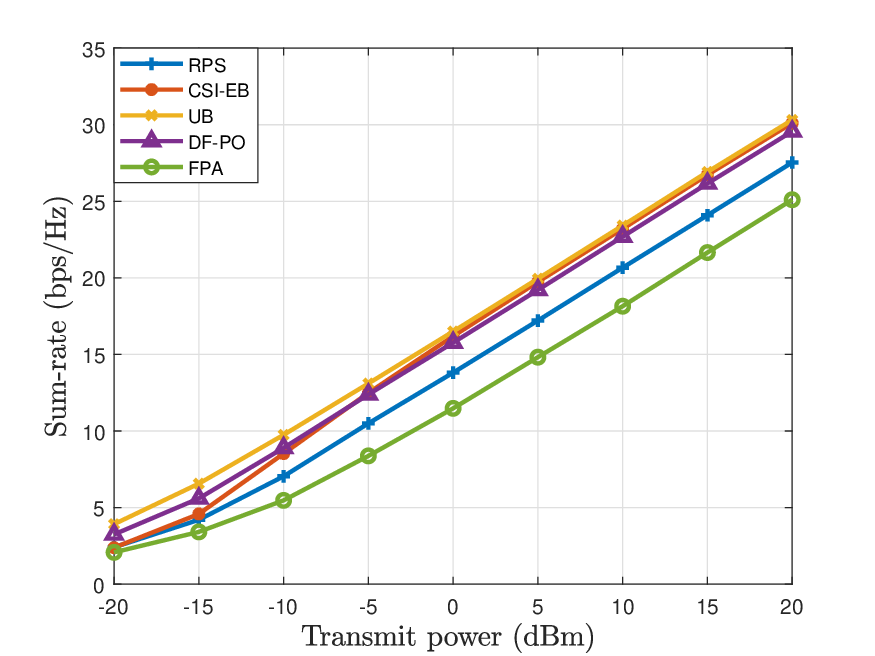}
	\caption{Sum-rates of respective methods versus the transmit power.}
	\label{fig:5}
\end{figure}

\subsection{Multi-User Scenarios}
We now consider the multi-user scenario. In our experiments, the number of users is set to $K=3$. For the multi-user case, the proposed DF-PO method requires a total number of $T_{all}=T(P_i+2P_z)$ pilot symbols for position optimization, where $T$ represents the number of pilot symbols used to estimate the channel matrix $\boldsymbol{H}$ at each position. In our simulations, we set $T=K+1$.  
Fig. \ref{fig:4} depicts the sum-rate of respective methods as a function of the number of pilot symbols, where the number of signal paths and the transmit power are respectively set to $L=70$ and $P_t=-5$dBm. From Fig. \ref{fig:4}, we see that the proposed DF-PO method achieves a substantial performance improvement over the competing algorithms in the regime where the number of pilot symbols is limited. The CSI-EB method is more favorable when the number of pilots is sufficiently large since the channel path parameters can be accurately estimated in such cases. This result demonstrates the sample efficiency of the proposed DF-PO method in solving the position optimization problem for MA-enabled systems. Such a merit is particularly advantageous for fast-changing propagation channels.


Fig. \ref{fig:5} plots the sum-rate of respective schemes versus the transmit power, where the number of pilots is set to $T_{all}=160$ and the number of signal paths is set to $L=70$. From Fig. \ref{fig:5}, we see that, given the same transmit power, MA-enabled systems can achieve a larger sum rate than the FPA system.
We also see that our proposed DF-PO method exhibits a better performance than the CSI-EB method when the transmit power is smaller than $0$ dBm. This indicates that the proposed method is robust against noise and is able to find an effective position even in the low SNR regime, while the CSI-EB method becomes ineffective when the SNR is low.

\begin{figure}[htb]
	\setlength{\abovedisplayskip}{1pt}
	\setlength{\belowdisplayskip}{1pt}
	\centering
	\includegraphics[width=9cm]{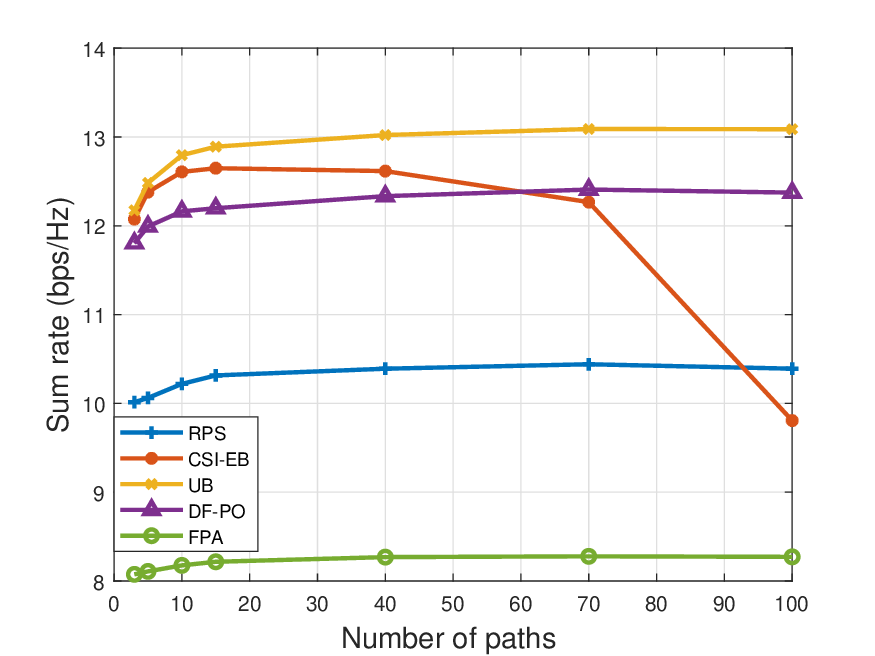}
	\caption{Sum-rates of respective methods versus the number of paths.}
	\label{fig:7}
\end{figure}

Fig. \ref{fig:7} show the sum-rates achieved by different methods as the number of signal paths varies, where the transmit power is set to $P_t=-5$ dBm and the number of pilot symbols is set to $T_{all}=160$. We see that our proposed method and the RPS method exhibit a consistent behavior as the number of paths increases. In contrast, the performance of the CSI-EB method is susceptible to the number of paths. This is because, as the number of paths increases, more channel parameters need to be estimated. When the number of pilots is insufficient to provide a reliable parameter estimation accuracy, the CSI-EB method would suffer a substantial performance loss. This result shows that the proposed DF-PO method can help find an effective position even for rich-scattering environments where the CSI-EB method usually fails.

\begin{figure*}[t!]
	\centering
	\begin{minipage}[b]{\linewidth}
		\subfigure[]{
			\includegraphics[width=0.5\linewidth]{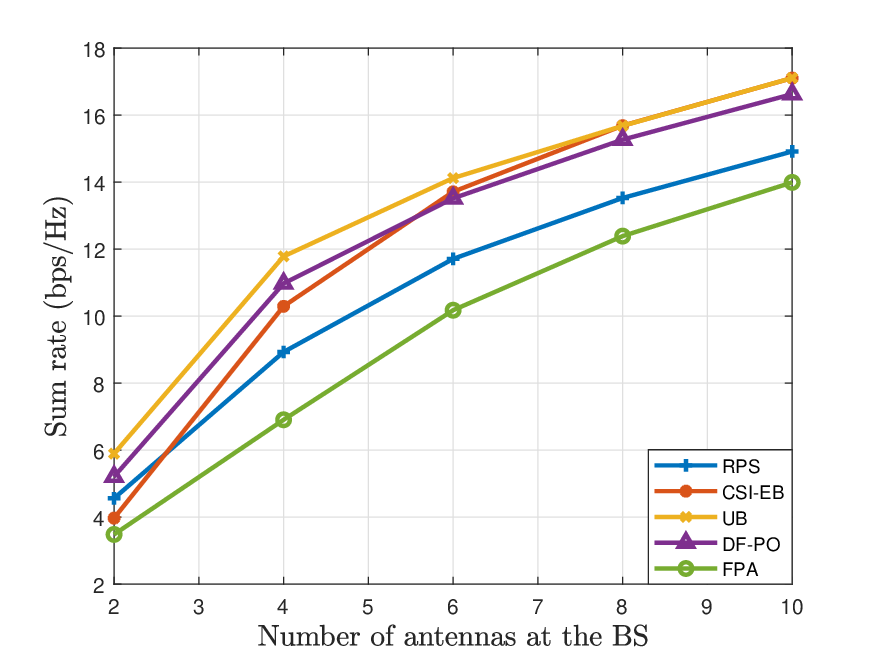}
		}
		\subfigure[]{
			\includegraphics[width=0.5\linewidth]{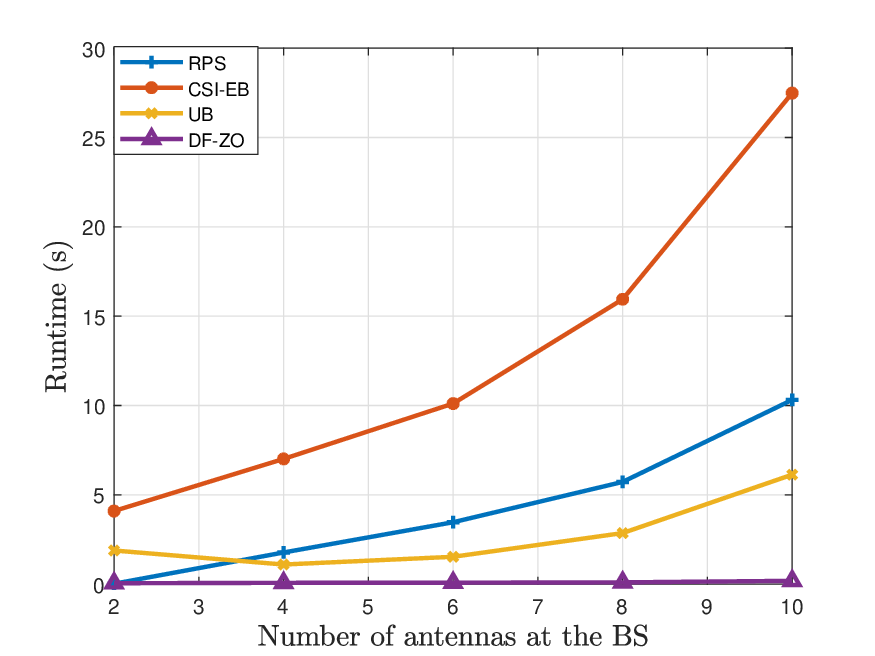}
		}
	\end{minipage}
	\caption{Sum-rates of respective methods versus the number of antennas.}
	\label{fig:8}
\end{figure*}

Finally, we examine the achievable rates and the average run times of respective methods as a function of the number of MAs at the BS in Fig. \ref{fig:8}, where the number of pilots and the number of paths are set to $T_{all}=160$ and $L=70$, respectively. From Fig. \ref{fig:8}(a), we see that, as expected, the performance of all methods improves as $M$ increases, as more antennas can provide a higher spectral efficiency. It is also noted that as the number of MAs increases, the CSI-EB method surpasses the proposed DF-PO method in terms of achievable rate. This is because more antennas brings more measurements and thus an improved estimation accuracy for channel path parameter estimation, which results in a better position optimization performance. Nevertheless, as the number of antennas grows, the average run time of the CSI-EB method undergoes a significant increase. In contrast, the computational complexity of the proposed DF-PO method remains almost constant as the number of antennas grows.

\section{Conclusion}\label{section6}
In this paper, we proposed a derivative-free method for position optimization for MA-enabled multi-user MISO systems, where the BS equipped with multiple MAs communicates with multiple single-antenna users. Different from existing position optimization methods that rely on global CSI over the entire movable regions, our proposed method treats the position optimization as a derivative-free optimization problem and adaptively refines the position based on the previous position measurements, without requiring global CSI. Simulation results demonstrated that the proposed method achieves a significant performance improvement (in terms of sample efficiency and computational efficiency) over the conventional position optimization methods that are based on channel estimation and reconstruction, particularly for challenging scenarios where the number of channel measurements is limited or the SNR is low.

\appendices
\section{Proof of Equation (\ref{Rate_MSE})}
\label{appendixA}
We have
\begin{align}
		R_k=\log_2\left(1+\frac{|\boldsymbol{w}^H_{k}\boldsymbol{h}_k(\boldsymbol{\vec{r}};\bm{\omega}_k)|^2}{\sum_{i=1, i\ne k}^{K}|\boldsymbol{w}^H_{k}\boldsymbol{h}_i(\boldsymbol{\vec{r}};\bm{\omega}_k)|^2+\|\boldsymbol{w}^H_{k}\|^2_2\sigma^2} \right)\label{Aeqn1}
\end{align}
Substituting (\ref{MMSE-combiner}) into (\ref{Aeqn1}), we arrive at
\begin{align}
	R_k=\log_2\left(1+\frac{|\boldsymbol{h}_{k}^H(\bm{J}^{-1})^H\boldsymbol{h}_k|^2}{\boldsymbol{h}_{k}^H(\bm{J}^{-1})^H 
		(\sum_{i=1, i\ne k}^{K}\boldsymbol{h}_i\boldsymbol{h}^H_i+{\sigma^2}\bm{I})\bm{J}^{-1}\boldsymbol{h}_{k}} \right) \label{Aeqn2}
\end{align}
where $\bm{J}\triangleq (\boldsymbol{H} \boldsymbol{H} ^H+{\sigma^2} \boldsymbol{I}_M) $ is a Hermitian matrix. We can further simplify (\ref{Aeqn2}) as
\begin{align}
	\begin{aligned}
	R_k&=\log_2\left(\frac{ \boldsymbol{h}_{k}^H \bm{J}^{-1} \boldsymbol{J} \bm{J}^{-1} \boldsymbol{h}_k }{
		\boldsymbol{h}_{k}^H(\bm{J}^{-1})\boldsymbol{J}(\bm{J}^{-1})\boldsymbol{h}_k
		- \boldsymbol{h}_{k}^H  \bm{J}^{-1}  \boldsymbol{h}_{k} \boldsymbol{h}_{k}^H \bm{J}^{-1}  \boldsymbol{h}_{k}} \right) 
		\\   &= \log_2\left(\frac{ \boldsymbol{h}_{k}^H\bm{J}^{-1}\boldsymbol{h}_k }{
			\boldsymbol{h}_{k}^H \bm{J}^{-1} \boldsymbol{h}_k
			- \boldsymbol{h}_{k}^H  \bm{J}^{-1}  \boldsymbol{h}_{k} \boldsymbol{h}_{k}^H \bm{J}^{-1}  \boldsymbol{h}_{k}} 
		 \right) 
		 	\\ & = \log_2\left(\frac{ 1}{1-\boldsymbol{h}_{k}^H \bm{J}^{-1} \boldsymbol{h}_{k}} 
		 	\right) 
		\end{aligned}\label{Aeqn3}
\end{align}
On the other hand, the MSE $e_k$ can be written as 
 \begin{align}
	\begin{aligned}
		&e_k = \mathbb{E}\Big[ |(\hat{ {s}}- {s})|^2\Big]  =\mathbb{E}\Big[ | \boldsymbol{w}_k^H\boldsymbol{y}_k- {s})|^2\Big] 
		\\&=\boldsymbol{w}_k^H\boldsymbol{H}\boldsymbol{H}^H\boldsymbol{w}_k-\boldsymbol{w}_k^H\boldsymbol{h}_k -\boldsymbol{h}_k^H \boldsymbol{w}_k+1+\sigma^2 \boldsymbol{w}_k^H\boldsymbol{w}_k
	\end{aligned}  \label{AMSE}
\end{align}
Substituting (\ref{MMSE-combiner}) into (\ref{AMSE}), we have 
\begin{align}
	\begin{aligned}
		 &e_k =\mathbb{E}\Big[ |(\hat{ {s}}- {s})|^2\Big] \\&=  \boldsymbol{w}_k^H\boldsymbol{H}\boldsymbol{H}^H\boldsymbol{w}_k-\boldsymbol{w}_k^H\boldsymbol{H} -\boldsymbol{H}^H \boldsymbol{w}_k+1+\sigma^2 \boldsymbol{w}_k^H\boldsymbol{w}_k
		\\&=\boldsymbol{h}_k^H \boldsymbol{J}^{-1} (\boldsymbol{HH}^H+\sigma^2)\boldsymbol{J}^{-1}\boldsymbol{h}_k-2\boldsymbol{h}_k^H \boldsymbol{J}^{-1}\boldsymbol{h}_k+1
		\\&=\boldsymbol{h}_k^H \boldsymbol{J}^{-1} \boldsymbol{J} \boldsymbol{J}^{-1}\boldsymbol{h}_k-2\boldsymbol{h}_k^H \boldsymbol{J}^{-1}\boldsymbol{h}_k+1
		\\&=1-\boldsymbol{h}_k^H \boldsymbol{J}^{-1}\boldsymbol{h}_k
	\end{aligned}   \label{Aeqn4} 
\end{align}
Thus we have
	\begin{align}
		R_k=\log\left( \left(e_k\right)^{-1}\right) 
	\end{align}
This completes the proof.

\section{Proof of Equation (\ref{P3})}
\label{appendixB}
Based on (\ref{Aeqn4}), the sum-MSE can be rewritten as 
	\begin{align}
	\begin{aligned}
		\sum_{k=1}^K{e_k}&=Tr(\boldsymbol{I}_k+\boldsymbol{H}^H\boldsymbol{J}^{-1}\boldsymbol{H})
		\\&=	Tr\left(\boldsymbol{I}_K\right)-Tr\left(\boldsymbol{J}^{-1}\boldsymbol{HH}^H\right) 
		\\&=Tr\left(\boldsymbol{I}_K\right)-Tr(\boldsymbol{J}^{-1}(\boldsymbol{J}-\sigma^2 \boldsymbol{I}_M))\\
		&=Tr\left({\sigma^2}\boldsymbol{J}^{-1}\right)+Tr\left(\boldsymbol{I}_K\right)-Tr\left(\boldsymbol{I}_M\right)\\
		&= {\sigma^2}Tr\left((\boldsymbol{H}\boldsymbol{H}^H+{\sigma^2} \boldsymbol{I}_M)^{-1}\right)+K-M\\
		&\overset{(b)}{=}{\sigma^2}Tr\left((\boldsymbol{H}^H\boldsymbol{H}+\frac{\sigma^2}{p} \boldsymbol{I}_K)^{-1}\right)	
	\end{aligned}\label{Aeqn5}
\end{align}

To establish the validity of (a), we employ the singular value decomposition (SVD) $\boldsymbol{H}=\bm{U}\bm{\Lambda}\bm{V}^H$,   where  $\bm{U}\in \mathbb{C}^{M\times M}$ and $\bm{V}\in \mathbb{C}^{K\times K}$  are unitary matrices, and $\bm{\Lambda}\in \mathbb{C}^{M\times K}$ contains singular values with $rank(\boldsymbol{H})\leqslant \min (M,K) $. Given  $K\leqslant M$, the channel matrix $\boldsymbol{H}$  has at most $K$ non-zero singular values ordered as  $\delta_1 \geq \delta_2 \geq \cdots \geq \delta_K\geq 0$. The proof proceeds through the essential equality 
\begin{align}
	\begin{aligned}
		(a)&={\sigma^2} Tr ((\bm{U}\bm{\Lambda}\bm{V}^H\bm{V}\bm{\Lambda}^H\bm{U}^H+{\sigma^2} \boldsymbol{I}_M)^{-1} )+K-M\\
		&={\sigma^2}Tr((\bm{U}\boldsymbol{\varOmega }_1\bm{U}^H+{\sigma^2}\boldsymbol{I}_M)^{-1})+K-M\\
		&={\sigma^2} Tr((\bm{U}\boldsymbol{\varOmega }_2\bm{U}^H+\sigma^2 \boldsymbol{I}_K)^{-1})\\
		&{=}{\sigma^2} Tr\left((\boldsymbol{H}^H\boldsymbol{H}+{\sigma^2}\boldsymbol{I}_K)^{-1}\right)
	\end{aligned}\label{Aeqn6}
\end{align}
where $\boldsymbol{\varOmega }_1=\bm{\Lambda} \bm{\Lambda}^H=\mathrm{diag}\{ \delta_1^2,\cdots ,\delta_K^2,\underset{M-K}{\underbrace{0,\cdots ,0}} \}$ and $\boldsymbol{\varOmega }_2=\bm{\Lambda}^H \bm{\Lambda}=\mathrm{diag}\{ \delta_1^2,\cdots ,\delta_K^2 \}$. This sequence of transformations confirms the equivalence required for (a), completing the proof. 

\bibliographystyle{IEEEtran}
\bibliography{refs}

\end{document}